\documentclass[aps,prl,reprint,superscriptaddress,nofootinbib]{revtex4-2}

\usepackage[utf8]{inputenc}
\usepackage{xcolor}
\usepackage{graphicx} 
\usepackage{epsfig}
\usepackage{epstopdf}
\usepackage{braket}
\usepackage[normalem]{ulem}
\usepackage[colorlinks]{hyperref}
\usepackage{bm}
\usepackage{siunitx}
\usepackage{amsmath}
\usepackage{amssymb}

\DeclareSIUnit\gauss{G}
\usepackage{soul}

\begin{document}

\title{A photon-interfaced ten qubit quantum network node}

\author{M.~Canteri}
\affiliation{Institut f\"ur Experimentalphysik, Universit\"at Innsbruck, Technikerstrasse 25, 6020 Innsbruck, Austria}

\author{\textcolor{black}{Z.~X.~Koong}}
\affiliation{Institut f\"ur Experimentalphysik, Universit\"at Innsbruck, Technikerstrasse 25, 6020 Innsbruck, Austria}

\author{\textcolor{black}{J.~Bate}}
\affiliation{Institut f\"ur Experimentalphysik, Universit\"at Innsbruck, Technikerstrasse 25, 6020 Innsbruck, Austria}

\author{\textcolor{black}{A.~Winkler}}
\affiliation{Institut f\"ur Experimentalphysik, Universit\"at Innsbruck, Technikerstrasse 25, 6020 Innsbruck, Austria}

\author{\textcolor{black}{V.~Krutyanskiy}}
\affiliation{Institut f\"ur Experimentalphysik, Universit\"at Innsbruck, Technikerstrasse 25, 6020 Innsbruck, Austria}

\author{B.~P.~Lanyon}
\email[Correspondence should be sent to ]{ben.lanyon@uibk.ac.at}
\affiliation{Institut f\"ur Experimentalphysik, Universit\"at Innsbruck, Technikerstrasse 25, 6020 Innsbruck, Austria}

\date{\today}

\begin{abstract}

We entangle each individual matter-qubit in a register of ten to a separate travelling photon. The qubits are encoded in a string of cotrapped atomic ions.  
By switching the trap confinement, ions are brought one at a time into the waist of an optical cavity and emit a photon via a laser-driven cavity-mediated Raman transition. The result is a train of photonic-qubits, each near-maximally entangled by their polarisation with a different ion-qubit in the string.  An average ion-photon Bell state fidelity of 92(1)\% is achieved, for an average probability for detecting each single photon of 9.1(8)\%. 
The technique is directly scalable to larger ion-qubit registers and opens up the near-term possibility of entangling distributed networks of trapped-ion quantum processors, sensing arrays and clocks.

\end{abstract}

\maketitle

There is a current research effort to develop the building blocks of future quantum networks \cite{Kimble2008,Wehnereaam9288, Wei22}: distributed matter-based nodes, for the storage and processing of quantum information, that are interconnected with photonic links, for the establishment of internode entanglement. 
Such networks have a broad range of envisioned applications \cite{Wehnereaam9288}: over distances of a few meters, they offer a path to scale up the qubit number in quantum computers and quantum simulators \cite{PhysRevA.89.022317, caleffi2022distributed}, over larger distances they offer a fundamentally new platform for secure communication \cite{Pirandola:20}, distributed sensing \cite{PhysRevLett.120.080501, PhysRevResearch.2.023052}, and timekeeping \cite{Komar2014}. Experiments have demonstrated the entanglement of two remote matter-based qubits via photon transfer, including qubits encoded in ionised atoms \cite{Moehring2007, Stephenson2020, Krut2022}, neutral atoms \cite{Ritter2012, Hofmann72, vanLeent2022}, solid state defect centres \cite{Bernien13, Hensen2015, Knaut2024}, quantum dots \cite{Delteil2016, Stockill2017}, and supercondcuting circuits \cite{Magnard2020}. Entanglement of three remote NV-centre qubits was achieved \cite{Pompili2021}.

In order to network intermediate-scale quantum processors, methods are first needed to connect the quantum states of multiple qubits in the processors to those of photons. Those photons can then be used to establish multiqubit entanglement between the processors, providing access to the full computational power of their joint Hilbert space. 
A few experiments have demonstrated coupling each individual in a register of two \cite{Inlek2017, Drmota2022, Kalb928} and three \cite{101kmmultimode} matter-qubits to separate photons. Two entangled Bell pairs between two remote NV-centre nodes have been established \cite{Kalb928}.

 \begin{figure}[t]
	\begin{center}
        \includegraphics[width=1\columnwidth]{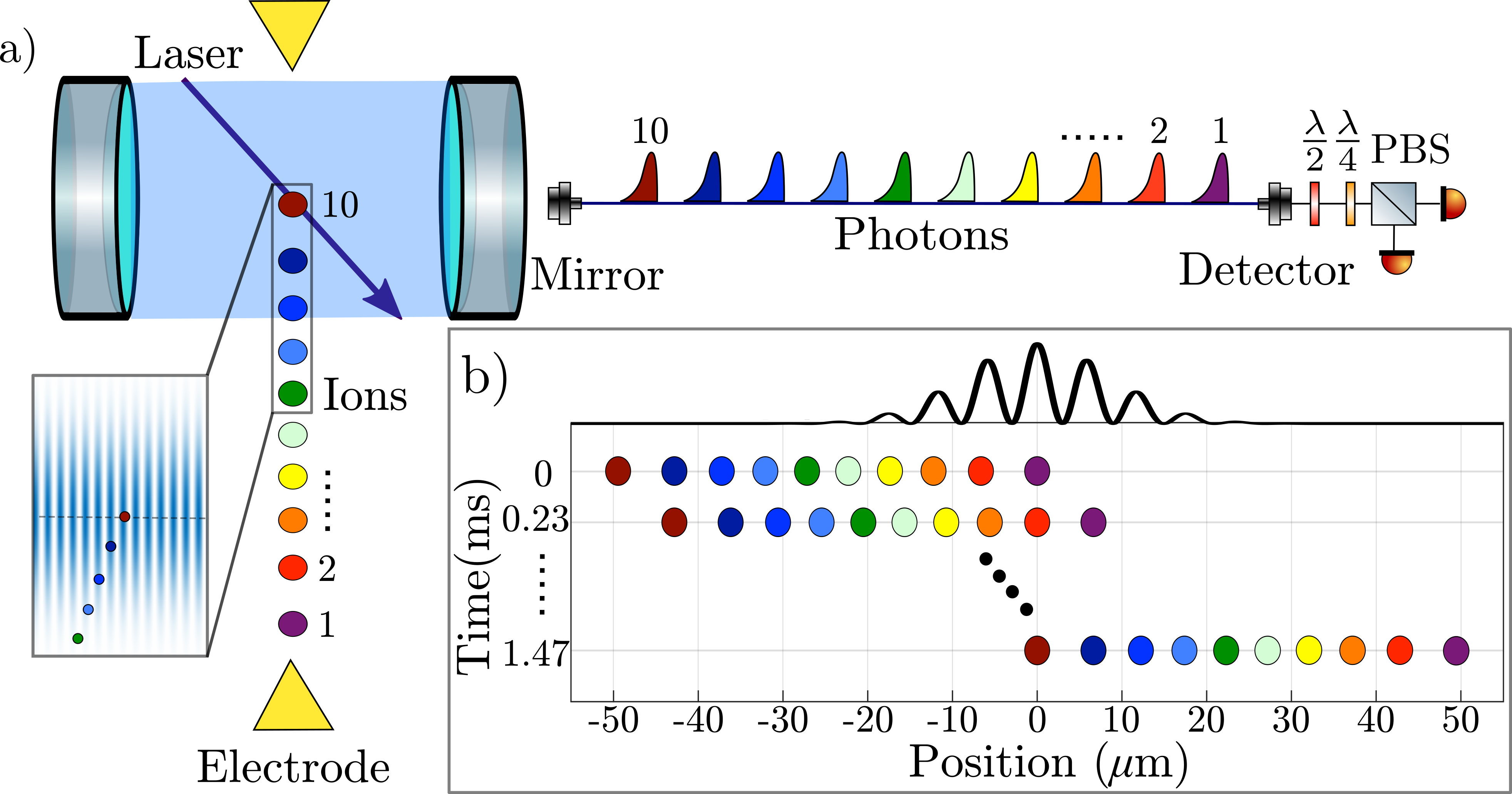}
		\caption{\textbf{Photon-interface for a multi-qubit register}. a) Concept image. Each photon is emitted by, and is polarisation-entangled with, a matching-coloured co-trapped ion. Configuration is shown after ion-string transport: a single-ion-focused laser beam just generated the last (10th, brown) photon from the (10th, brown) ion. Polarization analysis consists of half ($\lambda/2$) and quarter ($\lambda/4$) wave plates, and a polarizing beam splitter (PBS).
		Inset: zoom revealing vacuum cavity standing waves and 85.9$^{\circ}$ angle between string and cavity axis. b) The ion string is displaced stepwise through the cavity, bringing ions one at a time into the centre of the cavity waist and a standing wave maximum (Position \SI{0}{\micro\meter}). Black oscillating line: vacuum cavity standing wave (electric field squared) projected along the ion string axis.}
    		\label{fig:concept}
		\vspace{-9mm}
	\end{center}
\end{figure}

In this work, we demonstrate a scalable technique to entangle each matter-qubit in a register to a different propagating photon.
Qubits are encoded into a string of electrically trapped atomic ions. By modifying the trapping voltages, each ion in the string is brought sequentially into a photon-interfacing zone, corresponding to the focus of an optical cavity, and emits one photon. A train of photons is ideally obtained: each photon in the train is entangled with a qubit encoded in the ion that emitted it. In future, those photons are then available to entangle e.g., remote copies of the ion-register via photonic entanglement swapping, as demonstrated using one ion-qubit in our system over \SI{230}{\meter} \cite{Krut2022}. 
Entangling remote registers of trapped-ion qubits is a promising path to realise quantum networks and their most advanced applications. For example, strings of tens of $^{40}$Ca$^+$ ions---the species used in the present work--- are some of todays most advanced intermediate-scale quantum computers \cite{10.1063/1.5088164, Postler:2022ep}, quantum simulators \cite{PhysRevA.105.052426, Joshi:2023rq} and quantum sensor arrays  \cite{Marciniak:2022yc} and can also function as optical atomic clocks \cite{Matsubara_2008, PhysRevLett.102.023002, Matsubara:12, 10.1063/1.5090583, Guggemos_2019}.

 \begin{figure*}[th]
	\begin{center}	
        \includegraphics[width=2\columnwidth]{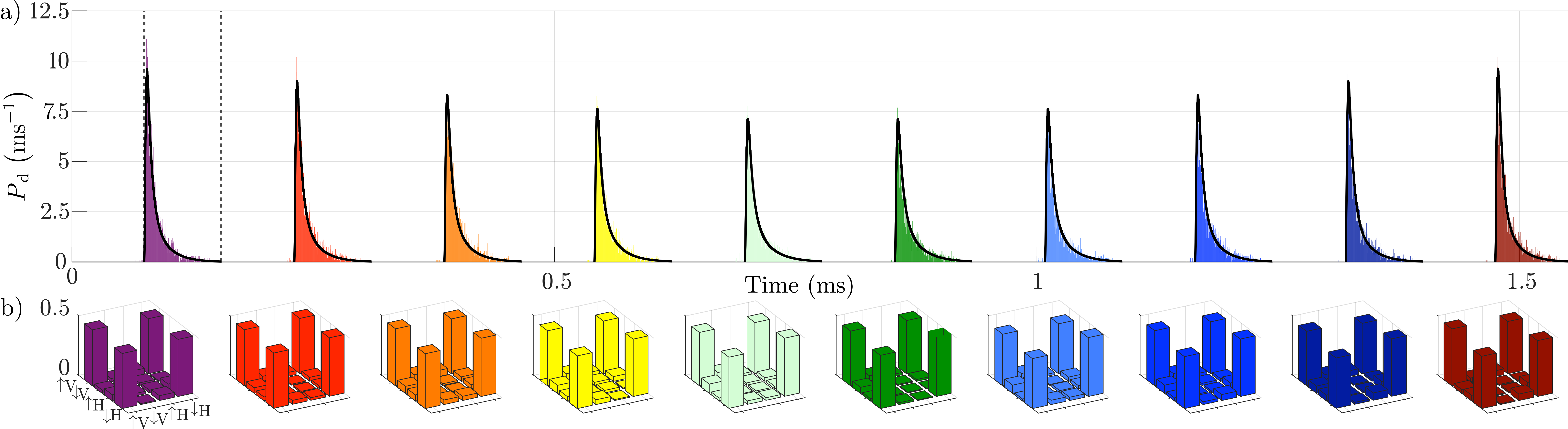}	
		\caption{a) Histograms of photon arrival times. Probability densities ($P_d$) are shown on the vertical axis: number of counts normalized by the number of attempts and by the \SI{0.2}{\micro\second} time-bin width. Ten single photon wavepackets are visible, color demarks the ion that is expected to have produced the photon (c.f. Figure 1). Solid black lines: numerical simulations. Dashed vertical lines show an example timewindow: counts within are associated with photon 1. b) Absolute values of the density matrices $\rho _{ii}^{Bell}$ of all ten ion-photon pairs. From left to right:  $\rho _{11}^{Bell}$, $\rho _{22}^{Bell}$ etc. Axis labels for the ion-qubits use the notation $\ket{\downarrow}=\ket{D}$ and $\ket{\uparrow}=\ket{D'}$. Absolute values of imaginary parts of any element in $\rho _{ii}^{Bell}$ are less than 0.06.}
    		\label{fig:wavepackets}
		\vspace{-5 mm}
	\end{center}
\end{figure*}

\twocolumngrid

Our experimental system consists of a string of ten $^{40}$Ca$^+$ atoms in a linear Paul trap with an integrated optical cavity for photon collection at \SI{854}{\nano\meter} (Figure 1a).
The cavity axis is close to perpendicular to the ion string axis, with a relative angle of \SI{85.9}{\degree}. 
The waist of the relevant vacuum mode of the cavity is \SI{12.31(8)}{\micro\meter} and the length of the ten ion string is \SI{49}{\micro\meter}, meaning that many ions lie far outside the cavity field (Figure 1b). By unbalancing the voltages on the endcap electrodes---which provide confinement along the string axis---different ions can be positioned at the centre of the cavity waist.

First, the ion at one end of the string (ion 1) is ideally placed at the optimal position for photon generation: a maximum of the vacuum standing wave at the centre of the cavity waist  (Figure 1).  A single photon is generated from ion 1 via a bichromatic cavity-mediated Raman transition (BCMRT) \cite{Stute2012, JosefPaper}, driven via a single-ion-focused  \SI{393}{\nano\meter} Raman laser beam with a \SI{1.2}{\mu }m intensity waist at the ion \cite{PhysRevLett.130.213601, 101kmmultimode}. 
A Raman laser pulse on the ion ideally generates the maximally-entangled state $\ket{\psi(\theta)}{=}(|D',V\rangle+e^{i\theta}|D,H\rangle)/\sqrt{2}$, where $|{D'}\rangle$ and $|{D}\rangle$ are the respective Zeeman states $|3^2{D}_{5/2}, m_j = -3/2\rangle$ and $|3^2{D}_{5/2}, m_j = -5/2\rangle$, and $|{V}\rangle$ and $|{H}\rangle$ are the vertical and horizontal polarization components of an \SI{854}{\nano\meter} photon emitted into the cavity vacuum mode, respectively. The angle $\theta$ is a phase set by the relative phase of the two frequency components in the bichromatic beam \cite{Stute2012}.

After the generated photon has exited the cavity, the ion string is moved (shuttled) along the string axis to a new equilibrium position in which the neighboring ion (ion 2) is now in the same optimal position for photon generation and an attempt to generate a photon (photon 2) is made.
This process---of bringing each ion one at a time into the position of optimal cavity coupling and attempting to make a photon---is repeated for all cotrapped ions, ideally yielding a train of single photons, each entangled with a different known ion-qubit in the string. 
On exiting the cavity through a preferred mirror, photons are collected into a single-mode optical fiber, which implements a rotation $U_p$ on the polarisation of each photon, and are then sent to a polarization analysis and detection setup (Figure 1a).

In detail, the experimental sequence consists of the following steps (with laser pulse durations given in brackets). First, the ions are initialised via Doppler cooling (\SI{8}{\milli\second}) followed by optical pumping into the state $\ket{S}=|4^2{S}_{1/2}, m_j = -1/2\rangle$ (\SI{40}{\micro\second}) and a wait time (\SI{60}{\micro\second}). Second, a Raman pulse (\SI{80}{\micro\second}) is applied to the ion at the cavity maximum (ion 1 in the first instance). 
Third, a change in the endcap electrode voltages is initiated that leads to a new ion string equilibrium position in which the neighbouring ion is at the cavity maximum. A \SI{60}{\micro\second} wait time now occurs to allow for ion transport \cite{SuppMat}. Steps two and three are then repeated until a Raman pulse has been applied to each ion in the string. 

In the last step, the ions are moved back to their initial position and ion-qubit measurement is carried out. For that, the $\ket{D'}$-state electron population of all ions is moved to the $\ket{S}$-state via a \SI{6.4}{\micro\second} $\pi$-pulse using a laser at \SI{729}{\nano\meter} on all ions. Next comes an optional \SI{729}{\nano\meter} $\pi/2$-pulse implemented on the $\ket{S}$ to $\ket{D}$ transition on all ions. The optional pulse is implemented when the ion-qubits are to be measured in the Pauli $\sigma_x$ or $\sigma_y$ basis: the optical phase of the pulse determines in which of the two bases the measurement is made. The optional pulse is not implemented when the ion-qubit is to be measured in the $\sigma_z$ basis. Finally, single-ion resolved qubit state detection is performed via electron shelving for \SI{5}{\milli\second} on all ions simultaneously, at which point the experimental sequence is concluded. The chosen ion-qubit and photon-qubit measurement bases are fixed throughout a single execution of the experimental pulse sequence. The sequence is repeated for all nine combinations of Pauli measurement bases for two qubits---set by different combinations of waveplate settings for the photon qubits and laser pulses for the ion qubits---allowing for tomographic reconstruction of all 100 possible combinations of two-qubit ion-photon pair density matrices $\rho_{ij}$ (ion $i$, photon $j$).

The pulse sequence was repeated 54000 times over 45 minutes, corresponding to the same number of attempts to generate a photon from each ion.  Figure 2a shows a histogram of all the single photon detection events, in which ten single photon wavepackets are clearly visible. The photons detected in e.g., the first (last) \SI{80}{\micro\second}-long time window are the ones expected to have been produced due to the corresponding Raman laser pulse applied to the first (tenth) ion, respectively. The total number of photon counts in the ten windows are $c_i=[5789,5388,5009,4740,4577,4441,4267,4689,5130,5869]$, 
and the corresponding photon detection probabilities ($P_i=c_i/54000$) are shown in Figure 3b. Only in 236 
of all 540,000 photon generation attempts was more than one photon detection event registered in the same time window. 

The data in Figures 2a and 3b show an efficiency profile: photons emitted by ions closer to the centre of the string were detected with lower probability. 
To predict this profile, a master-equation model of the photon generation process is developed that yields the expected photon efficiency and wavepacket shape on exiting the cavity, for each ion. 
In the model, the coupling strength of the photon generation process for the $i$th ion is proportional to $g\Omega_i^r$, where $g$ is the ion-cavity coupling strength and $\Omega_i^r$ is the reduced Raman laser Rabi frequency. The latter is given by, approximately, $\Omega_i^r=\Omega\prod_m(1-\eta^2_{i,m} \bar{n}_m)$, where $\Omega$ is the ground-state Rabi frequency on the $\ket{S_{1/2,-1/2}}$ to $\ket{P_{3/2,-3/2}}$ transition, $\eta_{i,m}$ is the Lamb-dicke factor for the $i$th ion in radial motional mode $m$ with mean thermal phonon number $\bar{n}_m$. 
$\Omega_i^r$ is reduced from $\Omega$ by the coupling of the Raman laser to motional sidebands of the ion string in the radial direction (the laser is perpendicular to the ion string and cavity axes). 
The reduction is greater for ions closer to the centre of the string because they participate more in the motion of the lower frequency modes, which have higher Lamb-dicke factors and higher $\bar{n}_m$ after Doppler cooling. 
Model parameters are set to values obtained from independent calibrations, except for the detection path efficiency $\xi$, which includes all photon losses after exiting the cavity up to and including photon detection. $\xi$ is obtained by fitting the model output to data in Figure 3b \cite{SuppMat}, yielding $\xi=0.36$. That value is lower than the upper limit determined from independent calibrations, which we attribute to the additional imperfections discussed below.

The measured single photon wavepackets and efficiencies are qualitatively well described by the predictions of the described master-equation model (Figures 2a and 3b).
The middle ions (5 and 6) are predicted to have a photon generation efficiency lower by a factor of 0.87 compared to the outer ions (1 and 10). 
This effect could be eliminated in future by ground state cooling the radial modes \cite{PhysRevA.93.053401}, after which the model predicts an average increase in photon generation efficiency per ion by a factor of 1.17.

Remaining differences between predicted and measured photon efficiencies are attributed to three additional sources of imperfection, beyond ion-string temperature. In the following, the reduction in photon generation efficiency contributed by these error sources is predicted by extensions to the model. 
The first additional error source are displacements $z$ of indivudal ions from the centre of the Raman laser beam focus, along the string axis, due to imperfect setting of endcap voltages. The position of each ion is estimated to have been set to within a displacement of $z=\pm$ \SI{0.2}{\micro\meter}. 
The model is extended to estimate the drop in photon generation efficiency per ion caused by such displacements, as a fraction of the value achieved without them \cite{SuppMat}. 
The extension considers the centre of the Raman beam to provide both the Rabi frequency for which the photon generation process (BCMRT) is resonantly driven and the maximum $g$ value. 
For $z=\pm$ \SI{0.2}{\micro\meter}, a photon generation efficiency drop by a factor of between 0.90 and 0.89 across the string is predicted, largely due to  
the photon generation process being driven off-resonantly for displaced ions.  
To reduce these fractional drops to less than a percent, the model predicts that ions should be positioned to within $\pm$ \SI{0.1}{\micro\meter} of the laser beam centre, which should be feasible by taking more data during ion-positioning to reduce statistical uncertainty.   

A second additional source of imperfections are coherent oscillations of the string's axial centre-of-mass (COM) mode, generated by the shuttling voltage steps. 
That oscillation is in a direction perpendicular to the tightly-focused Raman laser beam, yielding a time-dependent $\Omega$ and $g$, for the ion whose turn it is to generate a photon. 
Independent experiments with a single ion show that ion 10---shuttled the largest number of times before photon generation---could have an oscillation amplitude of up to $A_{com}=$ \SI{0.2}{\micro\meter}.
That amplitude is expected, by further model extensions \cite{SuppMat}, to cause a drop in photon generation efficiency for ion ten by a factor of 0.98, compared to the case $A_{com}=0$ and $z=0$. 
Combining e.g., $A_{com}=$ \SI{0.2}{\micro\meter} with a displacement of $z=\pm$ \SI{0.2}{\micro\meter} for ion ten is predicted to 
decrease this factor to 0.85. 
If not mitigated, shuttling-induced oscillations can increase for longer strings that contain more ions, which demand more voltage steps. 
Preliminary tests show that oscillation amplitudes can be significantly reduced by timing the shuttling steps with respect to the oscillation frequency such that e.g., every second step largely cancels the oscillation generated by the previous one. Moreover, we calculate that simply adding a low-pass electric filter with a cutoff at {\SI{40}{\kilo\hertz}} 
immediately before the endcap electrodes would reduce $A_{com}$ for a single voltage step by more than an order of magnitude, at the expense of increasing the shuttling wait time by a factor of maximally  1.1. Measurements with a single ion show no significant change in the motional state of the radial modes, at the level of a fraction of a phonon, due to the voltage changes applied in the ten ion experiment. 

The final error source considered is fluctuations in the endcap voltage difference at the local AC power frequency, causing ion-string oscillations along the string axis at \SI{50}{\hertz}. Voltage measurements yield an estimated ion amplitude of $A_{50}= $ \SI{0.1}{\micro\meter} at \SI{50}{\hertz}. Since the period of oscillation is much longer than that of a single Raman pulse, we can treat the effect as a static displacement. 
The effect on photon generation efficiencies will thus be less than that of a displacement $z=\pm$ \SI{0.1}{\micro\meter}. 
Further efforts to eliminate ground loops in the setup should significantly reduce $A_{50}$. 

From the data set, all two-qubit ion-photon density matrices $\rho_{ij}$ are tomographically reconstructed. The   concurrence $C$ \cite{Hill97_concurrence} quantifies the degree of entanglement in a two-qubit state, where $C=0$ is a non-entangled state and $C=1$ is a maximally entangled state achieved e.g., by $\ket{\psi(\theta)}$.  
The concurrences of all 100 ion-photon pairs (Figure 3a) show that all matched ion-photon pairs ($\rho_{ii}$) are strongly entangled, whilst all unmatched pairs ($\rho_{i, j\neq i}$) are not entangled: there is no evidence that any photon is entangled with any ion other than the one that was in the focus of the Raman laser when the photon was generated. 
Up to the precision of our measurements, ions not in the focus do not significantly participate in the BCMRT process such that they neither emit nor absorb cavity photons.

The states $\rho_{ii}$ are rotated with respect to each other by single qubit Pauli $\sigma_z$ rotations.  A numerical search is done to find the angles of these rotations. Specifically, the function $\sum_{m,n} F(U_z^m \rho_{mm}U_z^{m\dagger},U_z^n \rho_{nn}U_z^{n\dagger} )$ is maximized, where $U_z^k$ is a $\sigma_z$ rotation on ion-qubit $k$ (and identity on the photonic qubit) and $F$ is the mixed state fidelity function \cite{nielsen00}. 
The obtained rotation angles increase as a function of the ion number and are consistent with the effect of the independently-measured magnetic field gradient across the string of \SI{4.4(2)}{\gauss\per\meter} \cite{SuppMat}.
Two steps of manipulation bring the states $\rho_{ii}$  
into the Bell-state form ($\ket{\psi(0)}$). First,  the states $\rho_{ii}^{rot} = U_z^i \rho_{ii}U_z^{i,\dagger}$ are calculated. Second, the function $\sum_{m} F(U_p \rho_{mm}^{rot}U_p^{\dagger},\ket{\psi(0)} )$ is maximized over $U_p$: a single arbitrary one-qubit rotation of the photonic-qubits that is
attributed to the polarisation rotation applied by the optical fiber going to the photon analysis and detection station. 

The absolute values of the matrices $\rho _{ii}^{Bell}= U_p \rho_{ii}^{rot}U_p^{\dagger}$ 
are shown in Figure 2b. The Bell state fidelities  $\langle \psi(0)|\rho _{ii}^{Bell}|\psi(0)\rangle $, which range from 0.88$\pm$0.02 to 0.95$\pm$0.01, are shown in Figure 3c and compared with a model of ion-qubit decoherence. That model considers the coherence time of the $\ket{D}$-$\ket{D'}$ qubit of \SI{5.5(1)}{\milli\second}   \cite{SuppMat}, is applied to the last ion-photon state generated ($\rho _{10,10}^{Bell}$), and predicts a 0.02 drop in Bell fidelity between $\rho _{10,10}^{Bell}$ and $\rho _{1,1}^{Bell}$. 
A significant increase in coherence time is expected by mapping the ion-qubit states into the $\ket{S}$-state ground state manifold and performing spin-echos, as we demonstrated for two ions \cite{PhysRevLett.130.213601}. The highest Bell fidelities observed here are consistent with those observed for one and two ion strings in previous works \cite{PhysRevLett.130.213601, Krut2022}. We attribute errors not due to ion-qubit decoherence to a range of sources that each contribute infidelities at the level of $\leq 0.01$, including errors in \SI{729}{\nano\meter} laser pulses and polarisation analysis optics. A detailed study of the fidelity errors at that level is beyond the scope of this work.  \\

 \begin{figure}[t]
	\vspace{0mm}
	\begin{center}
        \includegraphics[width=1\columnwidth]{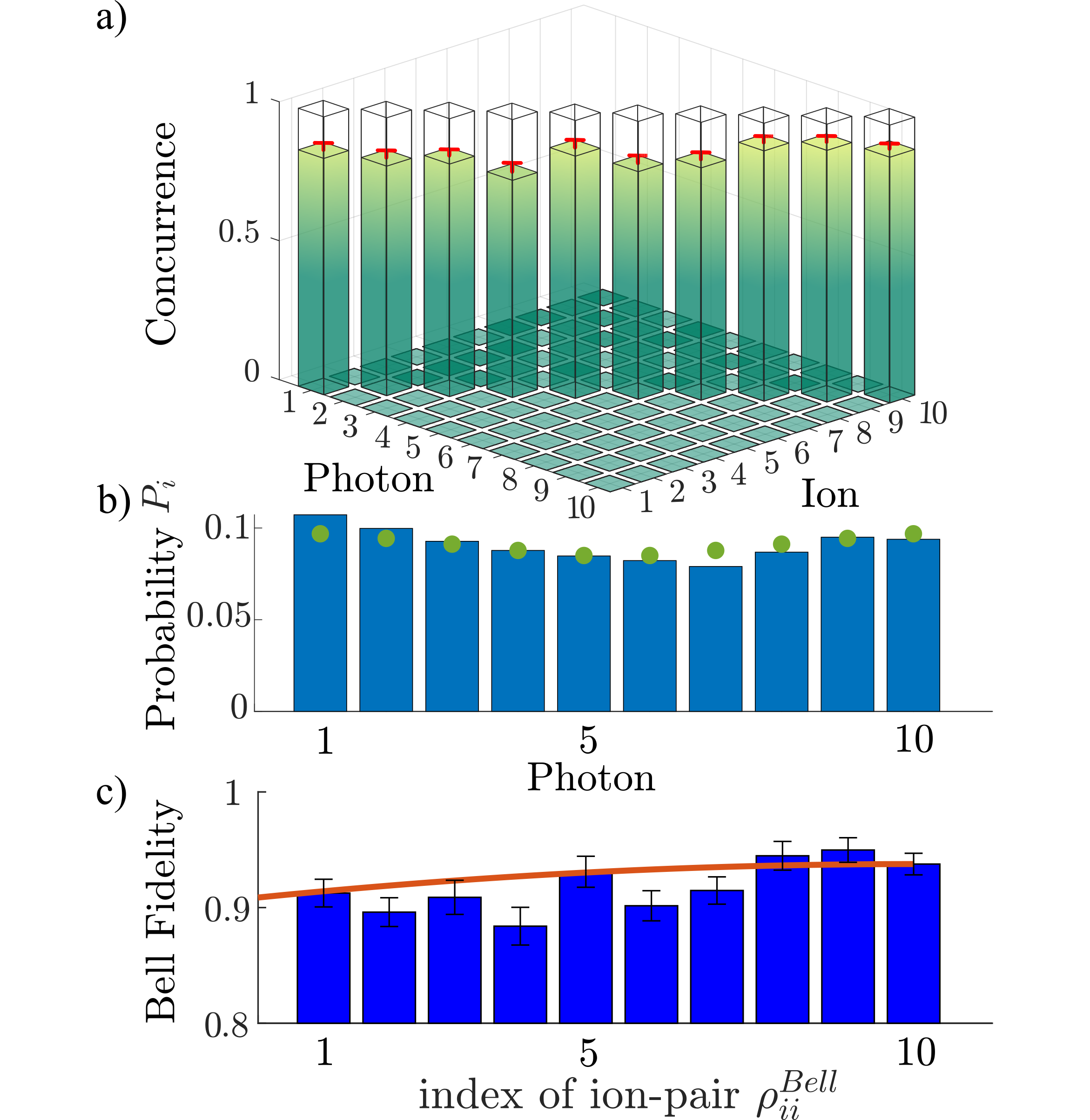}
		\vspace{-3mm}
		\caption{a) Concurrence of all 100 tomographically reconstructed ion-photon states $\rho_{ij}$. 	
		b) Photon detection efficiency. Blue bars: measured values $P_i$. Green points: model based on the finite temperature of the ion string.
c)  Blue bars with errors: Bell state fidelities of the ion-photon states $\rho _{ii}^{Bell}$ shown in Figure 2b. Orange curve: model based on calibrated ion-qubit decoherence. Errors bars in a) and c) are derived from Monte Carlo analysis of tomographic data. Error bars in b) are too small to be seen.}
    		\label{fig:results2}
		\vspace{-5mm}
	\end{center}
\end{figure}

A method was demonstrated that allowed each qubit in a ten-qubit register to be entangled with a separable fiber-coupled photon. 
The current formation rate of molecular ions in our system makes it impractical to work with strings containing more than around ten ions, probably due to trace water contamination. However, the presented method is otherwise directly scalable to strings containing many tens of ions that are routinely established in trapped ion systems, especially when combined with established methods for ground state cooling such ion-strings. 
Since ion transport was found to cause no significant change to the state of the radial motional modes \cite{SuppMat} it is expected that those modes can be readily used to implement high-fidelity laser-driven deterministic universal quantum logic gates on the qubit register \cite{Marciniak:2022yc}.

\begin{acknowledgments}
This work was funded in part by; the Austrian Science Fund (FWF) [Grant DOIs: 10.5576/Y849, 105576/P34066, 10.55776/F7116 and 10.55776/COE1]; the European Union under the DIGITAL-2021-QCI-01 Digital European Program under Project number No 101091642 and project name `QCI-CAT', and the European Union’s Horizon Europe research and innovation programme under grant agreement No. 101102140 and project name ‘QIA-Phase 1' and NextGenerationEU; the Österreichische Nationalstiftung für Forschung, Technologie und Entwicklung (AQUnet project). We acknowledge funding for B.P.L. by the CIFAR Quantum Information Science Program of Canada. The opinions expressed in this document reflect only the author’s view and reflects in no way the European Commission’s opinions. The European Commission is not responsible for any use that may be made of the information it contains. 
\end{acknowledgments}

%

\end{document}


\title{A photon-interfaced ten qubit quantum network node: Supplemental material}

\author{M.~Canteri}
\affiliation{Institut f\"ur Experimentalphysik, Universit\"at Innsbruck, Technikerstrasse 25, 6020 Innsbruck, Austria}

\author{\textcolor{black}{Z.~X.~Koong}}
\affiliation{Institut f\"ur Experimentalphysik, Universit\"at Innsbruck, Technikerstrasse 25, 6020 Innsbruck, Austria}

\author{\textcolor{black}{J.~Bate}}
\affiliation{Institut f\"ur Experimentalphysik, Universit\"at Innsbruck, Technikerstrasse 25, 6020 Innsbruck, Austria}

\author{\textcolor{black}{A.~Winkler}}
\affiliation{Institut f\"ur Experimentalphysik, Universit\"at Innsbruck, Technikerstrasse 25, 6020 Innsbruck, Austria}

\author{\textcolor{black}{V.~Krutyanskiy}}
\affiliation{Institut f\"ur Experimentalphysik, Universit\"at Innsbruck, Technikerstrasse 25, 6020 Innsbruck, Austria}

\author{B.~P.~Lanyon}
\email[Correspondence should be sent to ]{ben.lanyon@uibk.ac.at}
\affiliation{Institut f\"ur Experimentalphysik, Universit\"at Innsbruck, Technikerstrasse 25, 6020 Innsbruck, Austria}

\date{\today}

\maketitle

\tableofcontents

\section{Additional experimental details}

\subsection{Motional mode frequencies of the ten ion string} 
\label{modes}

The frequencies of all 20 motional modes of the string in the direction transverse to the ion string (radial modes) were measured via \SI{729}{\nano\meter} spectroscopy. Selected results are: $\omega_{20}=2\pi\times$\SI{2.0940(6)}{\mega\hertz} (highest COM mode); $\omega_{18}=2\pi\times$\SI{2.0469(6)}{\mega\hertz} (lowest COM mode); and $\omega_{1}=2\pi\times$\SI{1.1273(6)}{\mega\hertz} (lowest radial mode). The axial COM frequency was measured in a separate experiment for a three ion string in the same confinement, yielding $\omega_z=2\pi \times$\SI{358(9)}{\kilo\hertz}, where the uncertainty is set by the frequency resolution used for the spectroscopic scan.
Using these motional frequencies, we calculate the ion equilibrium positions (and thus separations) in the string by solving the classical equations of motion, yielding a string length of \SI{49}{\micro\meter}. 

\subsection{Shuttling electronics}\label{electronics}

For ion shuttling operations, voltages are established on the two endcap electrodes of the trap using the following components: 1. A commercial arbitrary waveform generator that defines and triggers the voltage waveform $V_\mathrm{in}(t)$. 2. A home-built voltage amplifier that produces two amplified, inverting and non-inverting outputs ($\pm 4V_\mathrm{in}(t)$). 3.  A high voltage supply to bias the two outputs $V_\mathrm{bias}$, resulting in a total voltage output of $V_\mathrm{bias} \pm 4V_\mathrm{in}(t)$. 4. A low-pass filter, with a designed bandwidth of \SI{100}{\kilo\hertz}. The low-pass filter is the last element before the two voltages are established at the endcap electrodes. One electrode has a voltage $V_\mathrm{bias} + 4V_\mathrm{in}(t)$ the other has $V_\mathrm{bias} - 4V_\mathrm{in}(t)$

The waveform $V_\mathrm{in}(t)$ corresponds to a sequence of ten steps in voltage between 10 different values $V_{\mathrm{in},j}$. The first nine steps move between those ten different voltages, which correspond to each ion being ideally located in the photon generation position. The tenth step returns to the initial voltage for ion readout and reinitialisation. In the ten ion experiment, a bias voltage of $V_\mathrm{bias}=$ \SI{160}{\volt} was used. The ten different voltage values set for $V_{\mathrm{in},j}$ were [4.450(1), 3.900(1), 3.400(1), 2.950(1), 2.550(1), 2.120(1), 1.700(1), 1.250(1), 0.750(1), 0.200(1)] volts, where the uncertainty in each value is given by the digital resolution of the waveform generator.

By applying a square pulse from the waveform generator, and monitoring the response on an oscilloscope, a rise time (10\% to 90\%) of $t_{rise}=$ \SI{9.0(5)}{\micro\second} was measured at the output of the voltage amplifier, corresponding to a 3dB bandwidth of  $0.35/t_{rise}=$ \SI{39}{\kilo\hertz}. 
Furthermore, fluctuations in the amplifier output voltages were measured for a minimal bias voltage. The difference between the two voltages revealed a peak-to-peak fluctuation of $\SI{85}{\milli\volt}$ at 50 Hz. Despite taking these measurements before the last low-pass filter, we expect the value to be reflected on the endcap electrodes as the passive filter doesn't add noise and its bandwidth is larger than the bandwidth of the voltage amplifier.

The $\SI{85}{\milli\volt}$ fluctuation is calculated to cause the ion string to oscillate along the string axis with an amplitude of \SI{123}{\nano\meter} at 50 Hz.
Here, the conversion factor \SI[per-mode = symbol]{2.899(7)}{\nano\meter\per\milli\volt} was used. That value was obtained in the following way. First, the calculated ion separations (see \ref{modes}) were plotted versus the voltages $4V_\mathrm{in,j}$ used in the experiment that ideally bring each ion into the position that is optimal for photon generation. Second, a linear fit yields the presented conversion factor as the gradient. The quoted uncertainty is derived from the uncertainty in the fit parameter. 

\subsection{Time allowed for shuttling}
\label{shuttlingtime}

The \SI{39}{\kilo\hertz} bandwidth of the electronics providing the voltage to the endcap electrodes results in an exponential relaxation of the voltage to the target value after the change of voltage $\Delta V = 4V_\mathrm{in,j}-4V_\mathrm{in,j-1}$ is produced for the ion shuttling. 
The measured voltage rise time of \SI{9.0(5)}{\micro\second} (10\% to 90\% amplitude) corresponds to the characteristic exponent time of \SI{4.1(2)}{\micro\second}.
After the onset of the Raman pulse (\SI{60}{\micro\second} after a voltage change is initiated) the remaining change of voltage yet to occur $\delta V$ is thus expected to be $\delta V = 4.3(2)\cdot10^{-7}\times \Delta V$, in the absence of noise.

Measuring the coordinate of an ion in the direction of the trap axis as a function of the delay from the voltage change trigger ($\tau_{shuttl}$) shows that it follows linearly the difference between the voltages on the endcaps. 
As a result, the fraction of the ion string's displacement that is still to occur, following initiation of a voltage change, is the same   
as the fractional voltage change still to occur $\frac{\delta V}{\Delta V}$.
For example, when applying a voltage step to shuttle ions by \SI{4.6}{\micro\meter}, a change in an ion's position between $\tau_{shuttl}=$ \SI{20}{\micro\second} and $\tau_{shuttl}=$ \SI{40}{\micro\second} was observed to not exceed \SI{0.15}{\micro\meter} -- the precision limit in the measurement performed.
By extrapolation of the exponential relaxation of an ion's equilibrium position, the remaining distance that it still has to travel after $\tau_{shuttl}=$ \SI{60}{\micro\second} from the initiation of a voltage change is calculated to be sub-nanometer in all the discussed experiments, in the absence of voltage noise.

The measurement of an ion's position as a function of delay $\tau_{shuttl}$ discussed above was performed in an experiment involving a two-ion string confined with an axial COM frequency of \SI{582(1)}{\kilo\hertz}. 
To measure the position of an ion at a given delay, a \SI{2}{\micro\second}-long pulse of the focused 393-nm beam was used with controllable delay from the shuttling trigger. The laser was detuned a few 10~MHz away from the corresponding atomic transition to increase the spontaneous scattering rate (for photon generation a detuning of several hundred MHz is used). 
The probability of pumping the ion to the dark state manifold ($\ket{3^2{D}_{5/2}}$) was measured as a function of the position of the laser beam's focus in the direction of the trap axis.
The ion's position was obtained as the center of a Gaussian fit to the data. 
Since this measurement yields the ion position averaged over a \SI{2}{\micro\second}-long laser pulse, it is not capable of resolving any \SI{582}{\kilo\hertz}  coherent oscillations. 

\subsection{Prospects for using the radial motional modes for quantum logic without additional cooling}
\label{radials}

The motional modes in the direction transverse to the axis of one-dimensional ion strings (radial modes) have been used in many works to perform universal quantum logic gates \cite{PhysRevLett.97.050505, Debnath:2016nk, Wright:2019hr, PhysRevA.100.022332, PhysRevA.107.032617,PhysRevLett.129.240504, chen2023benchmarking, Postler:2022ep}. In many of those works, for example \cite{Postler:2022ep}, lasers that off-resonantly drive one of the two ground-state cooled radial COM modes realises an entangling two-qubit gate between arbitrary ion-qubit pairs in the string. 

Calibration experiments, involving a single ground-state cooled ion in our system at similar trapping frequencies as for ten ions, show that the shuttling sequence in this work does not change the radial COM motional state at the level of statistical precision in our measurements (a fraction of a phonon). Moreover, the ambient motional heating rate of the highest radial COM mode is vanishingly low on timescales of many tens of milliseconds (less than or equal to a fraction of a phonon). We therefore expect that, after initial ground state cooling, the radial COM modes in our ten ion string will be suitable for realising a universal gate set after photon generation via shuttling, without the need for additional cooling. 
\subsection{729 nm laser}

 In comparison to our previous relevant works (\cite{JosefPaper}, \cite{PhysRevLett.130.213601}, \cite{101kmmultimode}), we here use a different \SI{729}{\nano\meter} laser and stabilisation system. The new laser is a titanium-doped sapphire laser that is frequency-stabilised to an external optical reference cavity \cite{HeleneMasterThesis, ArminMasterThesis}: the same reference cavity used to frequency-stabilise the \SI{393}{\nano\meter} Raman laser and an \SI{806}{\nano\meter} laser (the latter is for length-stabilisation of the ion-cavity). The linewidth of the \SI{729}{\nano\meter} laser was measured to be \SI{4.2(17)}{\hertz} over a measurement time of \SI{5}{\second}. Drops in the transmitted intensity through the locking cavity on the order of 15\% are observed. These drops are tens of microseconds long and occur every few tens of milliseconds. The origin of the drops is unclear and they likely limit the quality of \SI{729}{\nano\meter} $\pi$ and $\pi/2$ pulses such that they prepare single qubit target states with fidelities on the order of 0.99. 

\subsection{Positioning the ions}

Here, the procedure used to find the ten voltages $4V_{\mathrm{in},j}$ (see \ref{electronics}), that are employed to position each of the ten ions for photon generation, is described. The procedure consists of the following steps. 

\begin{enumerate}

\item The experiment begins in the condition in which the ion string axis and the cavity axis intersect, and this intersection point lies within the waist of the cavity \cite{schuppthesis}.  Therefore, there is a point along the ion string axis at which an ion can be placed (by varying the voltage difference of the endcaps) such that it can experience the maximum ion-cavity coupling strength in space. This relative alignment is achieved by translating the cavity position, with respect to the ion trap, using in-vacuum x/y/z nanopositioning stages \cite{schuppthesis} and, to within the precision of our measurements, is found to be stable on time scales of at least months. 
The optimal position of the ion along the string axis for photon generation is the one that leads to the highest detected single photon efficiency and is expected to correspond to positioning the ion at the centre of the waist in both the axial and transverse directions of the cavity axis. 
\item On the day of the experiment, ten ions are loaded into the trap. Ion 1 is put into the expected position for optimal photon generation, which involves setting the first estimate for the endcap differential voltage $4V_{in,1}$. This voltage is generated by running the waveform generator in DC mode (which becomes relevant later). 
\item The Raman laser is then aligned with the position of ion 1. This is done by preparing the ion in the state $\ket{S_{1/2,m_j=-1/2}}$ and then finding the laser position that maximises the rate at which the laser pumps the ion into the $\ket{3^2{D}_{5/2}}$ manifold. In that experiment, the ion-cavity is unlocked (its length is not actively stablised). 
\item A maximum of the vacuum cavity standing wave along the cavity is aligned with the position of ion 1. This is done by driving the photon generation process (Raman process) in a regime in which photons are continuously generated and translating the cavity along its axis on the nanometer scale until a maximum detected photon count rate is obtained. 
\item The frequency of the Raman laser is adjusted until the Raman process is on resonance. This is done by driving the photon generation process in a pulsed regime in which up to one photon is generated per pulse. The single photon detection efficiency is measured as a function of the Raman laser frequency, yielding a spectroscopic peak. A Gaussian fit to the data yields the resonance frequency. We refer to this process as setting the Raman resonance. 
\item  We now turn to work with the whole string. The goal now is to find the voltages $4V_{in,i>1}$ that maximise the photon detection efficiency for each ion. This is done with the waveform generator still in DC mode, by measuring the single photon detection efficiency as a function of the differential voltage $4V_{in}$ applied to the endcaps. The data shows ten peaks at ten different voltages, the centres of which are extracted by Gaussian fits.
The obtained voltages serve only as estimates for the final values since we find that voltages set in DC mode of the waveform generator are not identical to those in shuttling mode and thus all ten voltages need to be re-optimised once running the waveform generator out of DC mode for ion shutling.
\item The waveform generator is switched into shuttling mode, in which the dynamic waveform $V_\mathrm{in}(t)$ is produced that steps the voltage between the estimated values obtained in DC mode, and ideally bring each ion sequentially into the position of optimal photon generation. 
The voltages are then adjusted to maximise the detected photon generation efficiency from each ion individually. In our implementation of this optimisation process we estimate that the maximum values were obtained to within a precision of $4V_\mathrm{in,j}=\pm$ \SI{0.08}{\volt}. 
\item Finally, the Raman resonance is re-optimised (frequency of the Raman laser), as described in step 5, but this time averaged over all ten ions. Specifically, the single photon detection efficiency, averaged over all ten ions in shuttling mode, is measured as a function of the Raman laser frequency, yielding a single spectroscopic peak. A Gaussian fit to the data yields the resonance frequency for which the ten-ion experiment is subsequently carried out.

\end{enumerate}

\subsection{Cavity parameters}

The optical cavity around the ions is formed by two mirrors with a radius of curvature $R_C = \SI{9.984(7)}{\mm}$, separated by a length $l = \SI{19.906(3)}{\mm}$ in a near-concentric regime. More details about the construction and initial characterisation of the cavity can be found in \cite{JosefPaper}.

After the resistively-heated calcium source was accidentally operated at almost double the usual current for several minutes, the finesse $\mathcal{F}$ of the cavity was measured to be \num{30(1)e3}. This compares to a finesse of \num{54(1)e3} for previous works (e.g., \cite{JosefPaper, PhysRevLett.130.213601, Krut2022, 101kmmultimode}). We associate this with an increase in total photon losses in the cavity $\mathcal{L} = 2\pi/\mathcal{F}$. The cavity linewidth $\kappa = \frac{c\mathcal{L}}{4l}$, where c is the vacuum speed of light, and the probability of a photon escaping from the cavity mode to the desired output mode $P_{esc} = \frac{T_2}{\mathcal{L}}$, are therefore expected to have changed from $\{\kappa, P_{esc}\} = \{(2\pi)\times70(2)~ \textrm{kHz}, 0.78(3)\}$ 
to $\{\kappa, P_{esc}\} = \{(2\pi)\times126(2)~ \textrm{kHz}, 0.43(3)\}$.  The latter values, for the lower finesse cavity, are used in the models of the photon generation process in this work. 
\subsection{Photon detection efficiency budget}
\label{eff_budget}

The model for photon generation presented in \ref{photonmodel} predicts probabilities for single photon emission from the cavity output mirror into free-space. 
As described in the main text, a total detection path efficiency factor $\xi$ is then included (where $0\leq\xi\leq 1$), which captures all losses that the photons 
subsequently face up to and including the imperfect detectors. 

A detailed efficiency budget list for the single photon detection path is now given, based on independent measurements. The beginning of each element in the following list gives the transmission probability associated with a distinct part of the detection path, beginning with losses experienced once the photons leave the cavity output mirror and ending in photon detection.

\begin{enumerate}
\item 0.96(1): transmission of free-space optical elements that are between the cavity output mirror and first fiber coupler (see $P_{el}$ in \cite{JosefPaper}).
\item 0.84(3): efficiency of coupling the photons into the first single mode fiber. The value was measured with classical laser light. 
\item 0.73(3) and 0.68(3): efficiencies of the two paths in the polarisation analysis board (transmission and reflection respectively) measured with classical light. 
\item 0.80(3) and 0.88(3): detection efficiency of the two  detectors for \SI{854}{\nano\meter} (transmission and reflection paths respectively), calibrated by comparison with a manufacturer-calibrated detector.
\end{enumerate}

The products of the above listed efficiencies are 0.47(3) and 0.48(3) for the two detection paths, yielding an average value of $\xi_{max}= 0.48(3)$. 
This value should be considered an upper bound for the detection path efficiency in the ten ion experiment.

\section{Analysis and modelling}

\subsection{Rotations of the ion-photon density matrices $\rho_{ii}$ due to the magnetic field gradient}
\label{gradient}

\begin{figure}[t]
	\begin{center}
        \includegraphics[width=1\columnwidth]{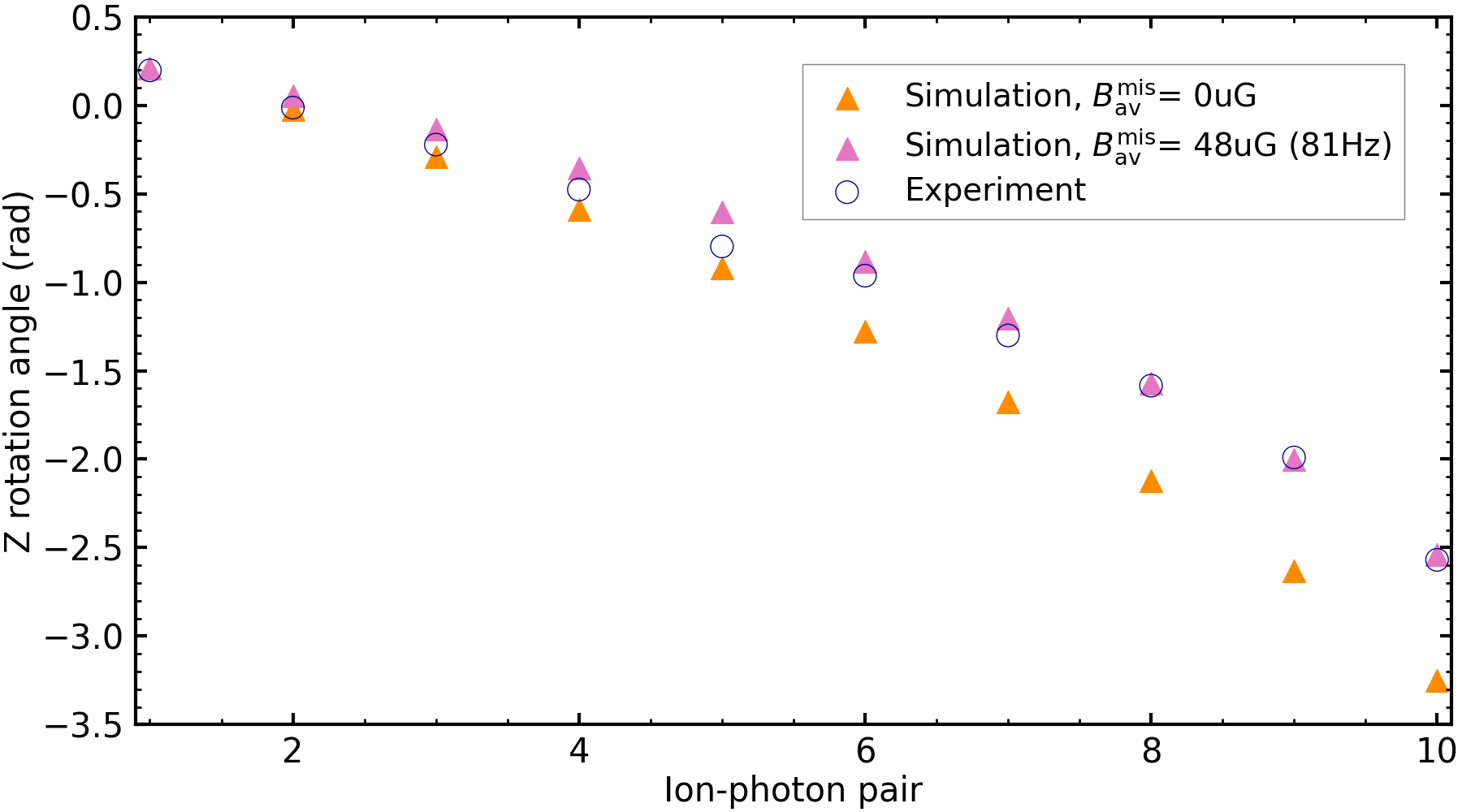}
		\caption{Ion-qubit Z rotations on the ion-photon states. Blue circles show angles found from tomography, obtained by maximising the sum of all pairwise fidelities (see main text). Orange points show simulation for $B^{\mathrm{mis}}_{\mathrm{av}} = 0$. Pink points show simulation for $B^{\mathrm{mis}}_{\mathrm{av}} = $ 48uG (81Hz). An overall phase offset of 0.32 and 0.24 respectively is added to the phases of the two simulations, to match the phase of ion-photon pair 1.}
    		\label{fig:shuttling_simulation}
		\vspace{-5mm}
	\end{center}
\end{figure}

As stated in the main text, ion photon pair density matrices $\rho_{ii}$ are found to be rotated with respect to each other by single qubit rotations on the ion qubits. These rotations are  captured by operators $U_z^i$: two-qubit unitary rotations corresponding to a $\sigma_z$ rotation on ion-qubit $i$ (and identity operator on photonic qubit $i$). To understand the origin of these rotations of the ion-qubits, an experiment was done to measure the magnetic field gradient $B_{\mathrm{grad}}$ in the trap. In this experiment, two ions were trapped with an axial center-of-mass (COM) mode frequency $\omega_{z} = 2\pi \times$ \SI{0.96(1)}{\mega\hertz} and ion-ion separation of \SI{5.74(4)}{\micro\meter}. Ramsey experiments using a \SI{729}{\nano\meter} laser on the $\ket{S}$ to $\ket{D'}$ transition were then performed with wait times of up to \SI{40}{\milli\second}.
The results yield a difference in transition frequency of $f_{\mathrm{diff}} = $ \SI{27(1)}{\hertz} when comparing the two ions. 
That frequency difference is consistent with a difference in magnetic field experienced by each ion of $B_{\mathrm{diff}} = $ \SI{25(1)}{\micro\gauss}, interacting with the ions through the linear Zeeman shift, corresponding to a magnetic field gradient along the string axis of $B_{\mathrm{grad}} = $ \SI{4.4(2)}{\gauss\per\meter}. 
\par
The shuttling procedure in this work aims to move the string to ten different locations. 
The simple model presented here considers the center of the ion string to perfectly follow the center of the trap ($X^{\mathrm{T}}$). 
$X^{\mathrm{T}}_n$ is the position of the center of the trap when the $n$-th ion is at the centre of the cavity waist. 
$X^{\mathrm{T}}$ is displaced sequentially from $X^{\mathrm{T}}_{n=1}$ to $X^{\mathrm{T}}_{n=10}$, and then returns to $X^{\mathrm{T}}_{n=1}$. 
$X^{\mathrm{T}}$ spends \SI{156}{\micro\second} at positions $X^{\mathrm{T}}_{n\neq 1}$, \SI{126}{\micro\second} at the initial position $X^{\mathrm{T}}_{n = 1}$, and \SI{30}{\micro\second} at the final position $X^{\mathrm{T}}_{n=1}$. 
Included in these times at each trap center position is the movement of $X^{\mathrm{T}}$ between each position $X^{\mathrm{T}}_{n}\rightarrow X^{\mathrm{T}}_{n+1}$, which is modelled by the spending \SI{30}{\micro\second} at $X^{\mathrm{T}}_{n}$ and \SI{30}{\micro\second} at $X^{\mathrm{T}}_{n+1}$. 
In the model, the ion at the cavity waist is prepared in the state $\frac{1}{\sqrt{2}}(\ket{D} + \ket{D'})$ at the time at which a Raman pulse would begin in the experiment.
At the $k$-th trap center position, the phase acquired by the $j$-th ion superposition state is calculated from the following expression: $2\pi t_k(f_{k, j}^{D\rightarrow D'} - (f_{L}^{S\rightarrow D} - f_{L}^{S\rightarrow D'}))$, where $f_{k, j}^{D\rightarrow D'}$ is the $D/D'$ transition frequency of the $j$-th ion superposition state at the $k$-th trap center position, $f_{L}^{S\rightarrow D}$ and $f_{L}^{S\rightarrow D'}$ are two laser frequencies used for subsequent analysis, and $t_k$ is the time spent at the $k$-th trap center position.
$f_{k, j}^{D\rightarrow D'} - (f_{L}^{S\rightarrow D} - f_{L}^{S\rightarrow D'}) \neq 0$ is expected to be caused by both a finite $B_{\mathrm{grad}}$ and a miscalibration of the average magnetic field over the 10 ions $B^{\mathrm{mis}}_{\mathrm{av}}$, via the linear Zeeman effect.
The accumulated phases are equivalent to the rotation angles of the $U_z^i$ operators. The rotation angles predicted from the model for $B^{\mathrm{mis}}_{\mathrm{av}} = 0$ are shown in Figure \ref{fig:shuttling_simulation} (orange points). 
$f_{L}^{S\rightarrow D}$ and $f_{L}^{S\rightarrow D'}$ are calculated from  a \SI{2}{\milli\second} long ramsey experiment which estimates $B^{\mathrm{mis}}_{\mathrm{av}}$. The pink triangles in Figure \ref{fig:shuttling_simulation} show the case for $B^{\mathrm{mis}}_{\mathrm{av}} = $ \SI{48}{\micro\gauss}. One can see the latter case is consistent with the measured rotations $U_z^i$.

\subsection{Model of the photon generation process}
\label{photonmodel}

Numerical simulations were performed to obtain estimations for the photon efficiencies and wavepackets. Specifically, the master-equation model of the laser-atom-cavity system is used from \cite{JosefPaper}.  The model parameters include the experimental geometries of the Raman laser, of the cavity and of the magnetic field, which are the same as described in \cite{PhysRevLett.130.213601} (in particular, see Sec. I.B and III.C of the supplementary material). The only other imperfection included in the model, besides those listed below, is a 10 kHz jitter of the resonance frequency of the cavity as described in \cite{JosefPaper}. \\

\noindent\emph{Temperature-}\\
The coupling strength of the photon generation process for ion $i$ is proportional to $g\Omega^r_i$, where $g=2\pi\times 1.53$ MHz is the maximum single-ion cavity coupling strength in our system and $\Omega^r_i$ is the Rabi frequency of the Raman laser on the relevant carrier transition from $\ket{4^2S_{1/2,-1/2}}$ to $\ket{4^2P_{3/2,-3/2}}$. Due to coupling between the Raman laser and the thermal motion of the ion string in the radial direction, $\Omega_i^r$ is approximately given by

\begin{equation}\Omega_i^r \approx \Omega\prod_m (1-\eta_{i,m}^2 \bar{n}_m), \label{eq:omega}\end{equation}

\noindent where $\Omega$ is the ground-state Rabi frequency of the Raman laser on the relevant transition, $\eta_{i,m}$ is the Lamb Dickie factor of the $m$-th motional radial mode and $\bar{n}_m$ is the mean  phonon number of that mode. There are a total of 20 radial modes of motion: two sets of ten and those sets oscillate in orthogonal directions. Equation 1 is valid in the Lamb-Dicke regime ($\eta{\ll}1$). For calculations the complete expression given in equation 3.11 of \cite{Roosthesis} is used. 
The Lamb-Dicke factor is given by

\begin{equation}\eta_{i,m}= \frac{2\pi}{\lambda} b_{i,m} \sqrt{\hbar/2M\omega_m}\cos(\theta_k), \label{eq:eta}\end{equation}

\noindent where $b_{i,m}$ are the mode eigenvectors, $M$ is the total mass of the string, $\omega_m$ are the mode frequencies, $\lambda$ is the wavelength of the Raman laser and $\theta_k=45^{\circ}$ is the angle between the k-vector of the Raman laser and either one of the two directions of radial motion.  

The model predicts the total probability $P_{i,c}$ for obtaining a photon in free space after exiting the cavity output mirror (and wavepacket temporal shapes). The total probability for detecting a photon is given by $\xi P_{i,c}$, where $0\leq\xi\leq1$ is a factor that accounts for all photon losses in the detection path after the photon exits the cavity (see section \ref{eff_budget}). Given the measured ten single-photon wavepackets, with individual photon detection probabilities $P_{i,d}$,  $\xi$ is determined by minimizing the function $\sum_i (P_{i,d} - \xi P_{i,c})^2$ over all photons $i$.

Now we turn to calibrate the parameters used in the model, in order to produce the theoretical predictions shown in Figures 2a and 3b of the main text. An estimate for $\Omega$ is obtained by measuring the AC Stark shift introduced by the Raman laser on the Raman transition, averaged over the ten ions. A detailed explanation of this method, for a single ion, is given in Appendix B of \cite{101kmmultimode} and references therein. 
The method involves measuring the photon detection probability as a function of the Raman laser detuning. To obtain an average AC Stark shift over the ten ion string, the total photon counts are measured over all ions, when implementing the shuttling protocol. 
We measured an AC Stark shift of the Raman transition of $2\pi\times$\SI{1.26(2)}{\mega\hertz}, which corresponds to a Rabi frequency of   $\Omega = 2\pi\times$\SI{53.2(4)}{\mega\hertz}. For a precise definition of this Rabi frequency see the supplementary material (Section III.C) of \cite{PhysRevLett.130.213601}.

Estimates for the mean phonon numbers ($\bar{n}_m$) are obtained via a calibration experiment with a single ion in the same trapping potential as for the ten ion experiment. After Doppler cooling, an average phonon number of $10\pm{2}$ was measured in each of the two radial COM modes. Those numbers were obtained by measuring Rabi flops on a \SI{729}{\nano\meter} transition and fitting them to a simple model within the Lamb-Dicke approximation. In the ten-ion model (Equation 1), these results are used in the following way: the average phonon number in the highest radial COM mode is set to 10 ($\bar{n}_{20}=10$). 
Then, we calculate the temperature $T_{10}$ for which the Bose-Einstein distribution results in the mean occupation number of 10 for a quantisation energy of $\hbar\omega_{20}$ (and zero chemical potential).  
The mean phonon numbers in each of the modes $m$ are set equal to the mean occupation numbers calculated for the Bose-Einstein distribution with the same temperature $T_{10}$ and energy $\hbar\omega_{m}$.
As a result, the lowest radial mode is predicted to have a mean phonon number of $\bar{n}_1=19$ and Lamb-Dicke factors of [0.0002,0.003, 0.016, 0.044, 0.071, 0.071, 0.044, 0.016, 0.003, 0.0002], for ions 1 to 10 respectively. In comparison, the highest radial mode has a mean phonon number of 10 and each ion has a Lamb-Dicke factor of 0.0028. One sees that the central ions couple more strongly (have the largest Lamb-Dicke factors) to the hotter (higher mean phonon number) modes.

Using the model as so far described, the theoretical predictions shown in Figures 2a and 3b of the main text were produced. A detection path efficiency of $\xi=0.36(3)$ was obtained via the fit previously described, where the uncertainty given is that of the cavity escape $P_{esc}=0.43(3)$. Note that the uncertainty in the fitted value for $\xi$ does not include the uncertainties in the other parameters that go into the model, e.g., $\bar{n}_{20}$.
That obtained value for $\xi$ is significantly lower than the upper bound of $\xi_{max}=0.48(3)$ from independent measurements of the detection path efficiency, described in section \ref{eff_budget} ($\xi/\xi_{max}=0.75(8)$). We attribute the majority of the difference to the other imperfections in the photon generation process that reduce its efficiency, in addition to ion-string temperature, that are described below. However, we cannot rule out that the detection path efficiency on the day of the ten-ion experiment was lower than $\xi_{max}$ due, for example, to a sub-optimal alignment of the optical setup for analysing photon polarisation. \\

\noindent\emph{Ion positioning errors (displacements)-}\\
The main text reports predicted photon generation efficiency drops due to anticipated displacements of ions from the centre of the Raman laser beam focus. To make those predictions, the model is extended to include the spatial profiles of the Raman laser Rabi frequency and the ion-cavity coupling strength along the trap axis $z$. The Rabi frequency is modelled as 
\begin{equation}
\label{omg_ex}
\Omega(z)= \Omega e^{\frac{-z^2}{\sigma^2}}
\end{equation}

\noindent where $\sigma = \SI{1.3(1)}{\micro\meter}$ is the waist of the Raman laser beam \cite{MarcoAmazingthesis}. The ion-cavity coupling strength is modelled as
\begin{equation}
\label{g_ex}
g(z) = g\cos(k z\sin(\phi))e^{-(z\cos(\phi)/\omega_0)^2}
\end{equation}

\noindent where $\omega_0= \SI{12.31(8)}{\micro\meter}$ is the waist of the cavity electric field and $\phi=4.1(1)^\circ$ 
is the angle between the ion-string and cavity axis, where both values are taken from \cite{schuppthesis}.

In the main text it is stated that the position of each ion is estimated to have been set to within approximately $\pm$\SI{0.2}{\micro\meter} of the centre of the Raman laser beam focus, along the string axis. Specifically, we estimate that the values of the voltages $4V_j$ were optimized to within  \SI{0.08}{\volt}. Using the conversion factor given in section \ref{electronics}, an error of \SI{0.08}{\volt} corresponds to a position displacement of $z=$ \SI{0.23(3)}{\micro\meter}. Using the extended model (equations \ref{omg_ex} and \ref{g_ex}) we estimate the drop in photon generation efficiency per ion caused by such positioning errors (displacements), as a fraction of the value achieved without them ($z=0$).
The model considers the centre of the Raman beam to provide both the Rabi frequency at which the photon generation process (Raman process) is resonantly driven and the maximum $g$ value. 
The results yield a predicted fractional drop of between 9\% and 11\% in photon generation efficiency across the string, where the central ions have the highest drop.  
These predicted efficiency drops are almost entirely caused by the photon generation process being driven off-resonantly, for ions positioned off centre (the effect due to the reduction in $g$ is minimal).

\begin{figure}[t]
	\vspace{5mm}
	\begin{center}
        \includegraphics[width=1\columnwidth]{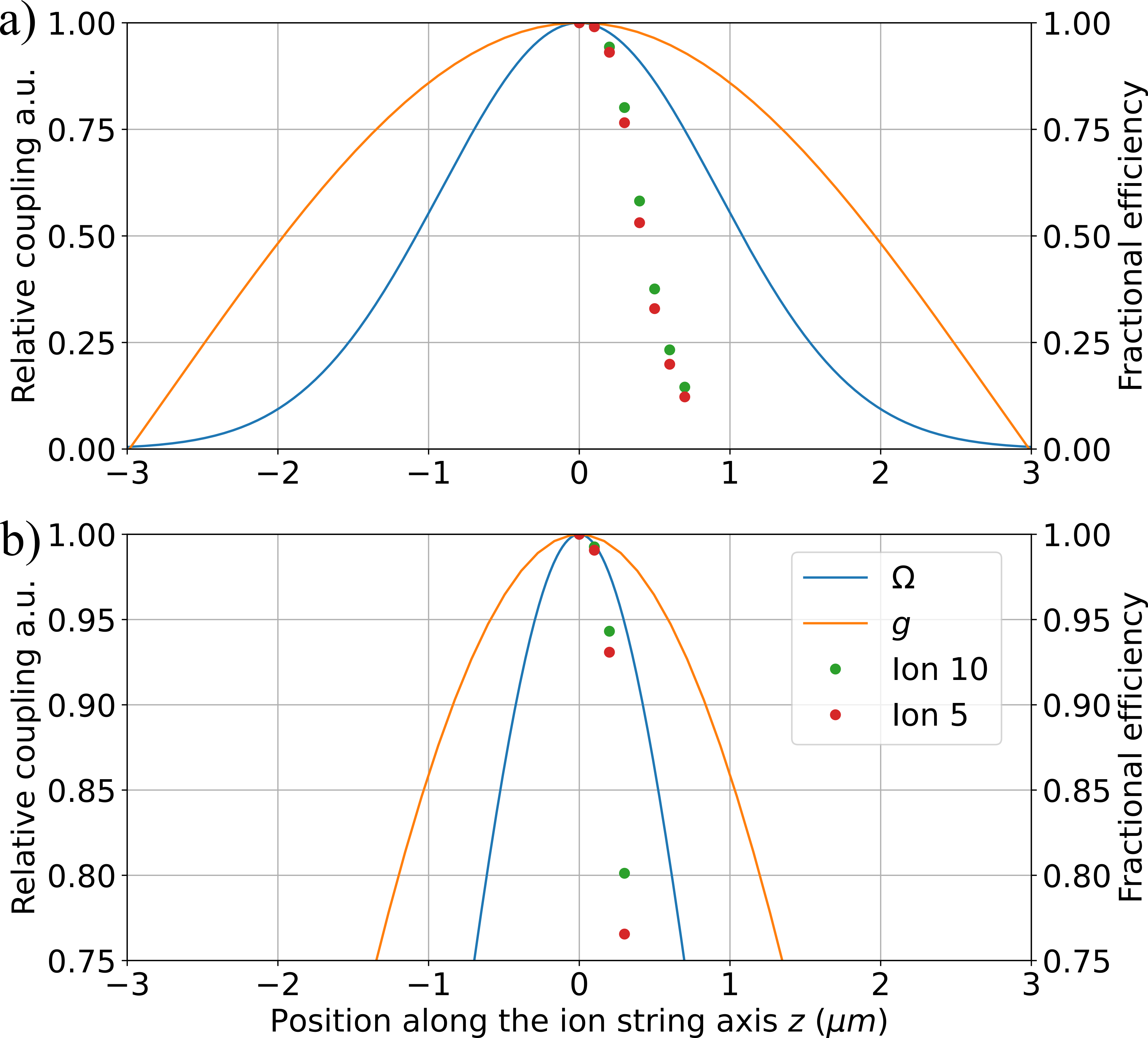}
		\caption{Prediction of how the photon generation efficiency drops as a function of ion displacement errors. Panel b) is a zoom in on panel a). The \emph{fractional} photon generation efficiency is plotted on the right hand vertical axis (the value for no displacement error is 1). The profiles of the cavity $g$ and laser coupling $\Omega$ are plotted together with simulations for ion's 5 and 10. In the model, the Raman resonance condition is satisfied at $z=0$.}
    		\label{fig:figure_profile}
		\vspace{-5mm}
	\end{center}
\end{figure}

Figure \ref{fig:figure_profile} shows how the photon generation efficiency from selected ions in the string is predicted to drop off as a function of displacements $z$ along the ion-string axis. 
As previously stated, the dominant cause of the efficiency drop is the Raman process becoming increasingly off-resonant for ions displaced from the laser beam centre: in the model the Raman process is set to be on resonant for the centre of the laser beam. 
The Raman process goes off resonance faster for ions nearer the centre of the string: increased thermal coupling with the Raman laser narrows the Raman resonance linewidth for those ions. 
In order to reduce the fractional efficiency drop to less than a percent, the model predicts that ions should be positioned to within $\pm$\SI{0.1}{\micro\meter} 
of the Raman laser beam centre, which should be feasible simply by devoting more time to calibration of the required voltages (and collecting more data) in order to reduce statistical uncertainty.  \\

\noindent\emph{Coherent oscillations-}\\
The main text refers to the generation of coherent oscillations of the string’s axial COM mode due to the shuttling steps. This motion causes the ion string to oscillate in a direction perpendicular to the focused Raman laser, yielding a time-dependent Rabi frequency and ion-cavity coupling strength, for the ion whos' turn it is to generate a photon. 
To model these oscillations, for any given ion, the model is extended to include a time dependent position of the ion under consideration, along the string axis of

 \begin{equation}
 z=A_{com}\sin(\omega_z t)+z_0,
 \label{osc_eqn}
 \end{equation}

\noindent where $A_{com}$ is the oscillation amplitude, $z_0$ is any additional displacement from the centre of the Raman beam focus (ion positioning errors) and $\omega_z$ 
is the axial COM angular frequency. The oscillations therefore lead to both a time dependent $\Omega$ and $g$ once inserted into Equations \ref{omg_ex} and \ref{g_ex}, respectively.   \\

In the main text it is written that calibration experiments with a single ion predict that the ten ion string could have a peak-to-peak oscillation amplitude of the axial COM mode of up to \SI{0.4}{\micro\meter} after the ninth shuttling step (which ideally brings the 10th ion into the focus of the Raman laser beam). We now provide more information about those single-ion calibration experiments.

\begin{itemize}
\item A single ion is loaded into the trap, with the same voltages set on the trap electrodes as for the ten ion experiment (and therefore ideally the same COM frequencies in all three spatial directions). 
\item Each experimental run then consists of variants of the following sequence. The ion is initialized via doppler cooling and optical pumping, then subjected to the same voltage waveform (dynamical sequence of endcap voltage changes), as carried out in the ten ion experiment. One of the voltage waveform parameters of importance here is the time $t_p$, which corresponds to the length of time between the onset of the voltage changes.
\item A \SI{0.5}{\micro\second}-long pulse from the Ramen laser is applied to the single ion at a time 8$t_p$+$t_d$+\SI{60}{\micro\second} following from the time the first voltage change is made, where \SI{60}{\micro\second} is the time allowed for the ion to reach an equilibrium position and $t_d$ is a positive tunable delay time that lies on the few microsecond scale. This laser pulse thus occurs when the voltage $4V_{\mathrm{in},10}$ has been established after the ninth change in voltage.
 \item In contrast to the Raman pulse used to generate photons, the \SI{0.5}{\micro\second}-long pulse here is set on resonance with the relevant $S-P$ transition, such that the dominant effect is to resonantly excite the electron to the $P_{3/2}$ manifold, from which the electron can decay to the $D_{5/2}$ manifold (the cavity plays no role in this experiment). The intensity of the laser is set such that the probability of finding the electron to be in the $D_{5/2}$ manifold afterwards $P_{\text{dark}}$, when the Raman beam focus is well centred on the ion's position, is around 0.8. $P_{\text{dark}}$ is estimated via electron shelving.
\item In the first experiment the values $t_p=$ \SI{156}{\micro\second} and $t_d=$ \SI{0}{\micro\second} are chosen and $P_{\text{dark}}$ is measured as a function of the position of the centre of the laser beam along the ion string axis: an acousto-optic deflector (AOD) is used to deflect the focus along this axis. The result is a Gaussian-shaped profile which is fit to the function $P_{\text{dark}}= P_0e^{-(f-f_0)^2/2\tau^2}$ where $f$ is the frequency of the AOD. The AOD frequency (laser focus position along the string axis) is extracted that corresponds to half way up one of the sides of the Gaussian. At this point, a change in the position of the ion along the string axis will result in a proportional change in $P_{\text{dark}}$, over a range of positions set by the relevant width $\tau$ of the Raman laser beam.
\item In the second experiment the laser position is fixed to the one corresponding to approximately half way up one of the sides of the Gaussian, and $P_{\text{dark}}$ is measured as a function of the wait time $t_d$. The data show oscillations in $P_{\text{dark}}$ as a function of $t_d$ with a frequency of $\omega_z$ and probability amplitude $A_{P_{\text{dark}}}$. The  value $A_{P_{\text{dark}}}$ is extracted by a sinusoidal fit function---$A_{P_{\text{dark}}}\sin (2\pi\omega_z t_d ) + P_{\text{offset}}$---and converted to an amplitude in distance along the string axis by 
inverting the Gaussian fit function for  $f(P_{\text{dark}})$  as $f(P_{\text{dark}}) = f_0+\sqrt{-2\tau^2\ln(P_{\text{dark}}/P_0)}$. The amplitude in frequency is then $A_f = \frac{1}{2}f(P_{\text{offset}}+ A_{P_{\text{dark}}}) + \frac{1}{2}f(P_{\text{offset}}-A_{P_{\text{dark}}})$. Finally, $A_f$ is converted to position on the trap axis via a previously calibrated factor of $c=$\SI[per-mode=symbol]{3.03}{\micro\meter\per\mega\hertz} \cite{MarcoAmazingthesis}, i.e.  $A_{com}=cA_f$.
\item The second experiment is repeated for a range of values $t_p$. 

\end{itemize}

The results show a periodic structure in $t_p$ with maximum values (and minimum values) of $A_{com}$ (Eqn. \ref{osc_eqn}) spaced by times consistent with integer multiples of the \SI{2.8}{\micro\second} period of the trap axial COM mode. Changing $t_p$ corresponds to changing the time between the kicks applied to the ion string by the voltage changes and therefore the extent to which they constructively or deconstructively excite the ion string's COM motion.  The largest $A_{com}$ measured in the single-ion experiment was \SI{0.2}{\micro\meter}, and was observed for $\tau_p=$ \SI{156.5}{\micro\second}. In the ten-ion experiment $\tau_p=$ \SI{156}{\micro\second}. However, we do not know that the axial COM frequencies in the two experiments were set the same to better than a few kHz (the relevant one ion and ten ion experiments were done several months apart), and therefore we take the largest measured amplitude with the single ion to represent the worst case scenario for the ten-ion experiment.

\subsection{Simulation of shuttling-induced coherent oscillations using classical equations of motion}

Shuttling can excite (generate coherent oscillations of) the axial centre of mass motion of the ion string, as considered in \ref{photonmodel}.  We model the motion $z(t)$ of the axial centre of mass of the $N$-ion string by numerically solving the classical equation of motion for a total charge $N\times e$ and total mass $M=N\times m$ ($e$ is the electron charge, $m$ is the single ion mass) that is confined in a 1D harmonic trap electric potential $U(z) = M\omega_z^2(z-z_0)^2/2$. 
The goal is to predict the amplitude of oscillation $A_{com}$ in equation \ref{osc_eqn} that is excited during the shuttling sequence. 
The change in confining potential during shuttling is modelled by setting $z_0 = c(t)$, where $c(t)$ is the expected time-dependent position of trap's centre, which is modelled to depend linearly on the difference of the voltages on the two endcap electrodes. 
For the sequence of nine voltage steps 
used in the experiment ($4V_{\mathrm{in},j}$), $c(t)$ is generated numerically by sequentially applying low pass filters to a function $s(t)$ consisting of a sequence of nine instantaneous steps with equal time intervals $t_p$ between them. The amplitude of each step $j$ is set proportional to the voltage $4V_{\mathrm{in},j}$ and normalized such that $s(9t_p)-s(0) =$ \SI{49}{\micro\meter} (the ten-ion string length).  
First a first-order low-pass filter is applied to $s(t)$, representing the limited bandwidth of the voltage amplifying board. 
Second, a second-order low-pass filter is applied, to ensure continuity of the time derivative $\dot{c}(t)$ as observed in the measured voltage $4V_{in}$ trace.
The cutoff values of those filters are set to \SI{35}{\kilo\hertz} and \SI{80}{\kilo\hertz} respectively, found by the best fit of the generated function $c(t)$ to the measured time trace of the voltage waveform as described in section \ref{electronics}.

The model predicts the following results.  First, the amplitude $A_{com}$ (Eqn. \ref{osc_eqn}) after the first shuttling step is calculated to be \SI{0.02}{\micro\meter}.
Second, after nine shuttling steps the maximum amplitude $A_{com}$ is calculated to be \SI{0.21}{\micro\meter}, and is achieved when the time intervals $t_p$ between the shuttling steps are set to an integer number of the trap period $2\pi/\omega_z$. 
This result is consistent with the maximum amplitude of \SI{0.21(4)}{\micro\meter}, measured for a single ion.
Third, we find that adding an extra first-order low pass filter (a third filter in the chain) with a cutoff of \SI{40}{\kilo\hertz} is predicted to result in a reduction of $A_{com}$, both for a single shuttling step and for the worst-case nine steps, by at least one order of magnitude.
Adding this third filter results in increasing the time to achieve the same equilibrium position after the shuttling by a factor of 1.1. 
This filter could be added either as the last element before the endcap voltages, e.g., by implementing it as hardware, or as the first element, by  e.g., programming it into the arbitrary waveform generator.

\subsection{Model of ion-qubit decoherence}
\label{decoherence}

This section presents the model of ion-qubit decoherence used to predict the drop in Bell state fidelity presented in Figure 3c (solid line) of the main text.
After emitting a photon, the ion-qubit state is stored in the $D$ and $D'$ states, where it experiences decoherence over time due to e.g., fluctuating magnetic fields. The qubit in the last ion to emit a photon is stored in the $D/D'$ states for the least amount of time before readout and therefore the ion-photon state $\rho_{10,10}$ is the least affected by ion-qubit decoherence. We model the effect of ion-qubit decoherence with a Kraus map $\mathcal{E}(\rho)$, yielding a prediction for how the state $\rho_{10,10}$ would decohere in the case of additional storage time $t$ in the $D/D'$ manifold. 
We used the following Kraus map
\begin{equation}
\mathcal{E}(\rho,t) = \frac{1}{2}(p(t)+1)\rho + \frac{1}{2}(1-p(t))\left[\sigma_z\otimes\mathbb{I}\right] \rho\left[\sigma_z^\dagger\otimes\mathbb{I}\right]
\end{equation}
where $p(t)$ can be interpreted as the probability for a $\sigma_z$ phase flip to happen on the ion-qubit. We model the probability as a Gaussian distribution 
\begin{equation}
p(t)=e^{-t^2/2\sigma^2}
\label{gauss}
\end{equation}

with the characteristic decay time $\sigma$. Finally, we calculate the Bell state fidelity $F(t) = \langle \psi(0)|\mathcal{E}(\rho_{10,10},t)|\psi(0)\rangle $ over the time of the experiment $t= $ \SI{1.4}{\milli\second}.

The decay time $\sigma$ was obtained in a measurement. Here, we measured the Ramsey contrast for a qubit encoded in the $D/D'$ states for three ions trapped in the same configuration as the 10 ion string. A fit of the function Eqn. \ref{gauss} to the data yields the decay $\sigma$ for each one of the three ions: the average value is $\sigma = $ \SI{5.5(1)}{\milli\second} and that value without error is used in simulations. No statistically-significant differences between the values for the three ions were observed

The model plotted in Figure 3c of the main text is the curve $F(\tau)$ where the time has been shifted as  $\tau = |t-1.4|$ ms since the first ion to emit a photon spends the most amount of time in the $D/D'$ state and therefore its qubit experiences the most decoherence.

\bibliography{biblio.bib}